# Empowering Large Scale Quantum Circuit Development: Effective Simulation of Sycamore Circuits

Venkateswaran Kasirajan[1], Torey Battelle[2], Bob Wold[1]

[1]Quantum Rings Inc., Broomfield, CO 80021, USA

[2]Arizona State University, Tempe, AZ, 85281, USA

Simulating quantum systems using classical computing equipment has been a significant research focus. This work demonstrates that circuits as large and complex as the random circuit sampling (RCS) circuits published as a part of Google's pioneering work [4-7] claiming quantum supremacy can be effectively simulated with high fidelity on classical systems commonly available to developers, using the universal quantum simulator included in the Quantum Rings SDK, making this advancement accessible to everyone. This study achieved an average linear cross-entropy benchmarking (XEB) score of 0.678, indicating a strong correlation with ideal quantum simulation and exceeding the XEB values currently reported for the same circuits today while completing circuit execution in a reasonable timeframe. This capability empowers researchers and developers to build, debug, and execute large-scale quantum circuits ahead of the general availability of low-error rate quantum computers and invent new quantum algorithms or deploy commercial-grade applications.

## I. INTRODUCTION

Quantum computing, the next significant leap in computing, is poised to revolutionize the way we solve complex problems in scientific and commercial applications. The potential of quantum computing is immense, and achieving quantum advantage, the point at which a quantum computer can out-perform a classical computer (improved accuracy and/or precision, faster and more efficient time-to-solution, even the resolution of previously intractable scenarios), is a crucial milestone in this field.

Determining a manifestation of quantum advantage is a matter of delicacy. It draws on all aspects of quantum computation: hardware, software, architectures, benchmarks, and analysis methods. Researchers have developed a wide array of approaches that aim to provide a robust methodology to evaluate the quality of quantum computation.

In a historic 2019 research paper [4], Google used random circuit sampling [5] as a tool to perform cross-entropy benchmarking (XEB) [5] of their quantum system and used the results to make a quantum supremacy claim. In this experiment, Google's quantum computer completed a random sampling task in 200 seconds, a feat that would have taken a state-of-the-art supercomputer approximately 10,000 years [4]. This significant leap in computational speed demonstrated the potential of quantum computing and marked a crucial milestone in the field. Since then, random circuits [3] have been extensively studied and are often used in benchmark studies [9-18].

Since Google's original work was published, classical computing capacities and software algorithms have significantly advanced, enabling the simulation of larger and more complex quantum circuits. This has raised the bar for the definition of quantum advantage, as seen in more recent publications [18], illustrating that "Quantum Advantage" will be a moving target.

The central thesis of this paper is to demonstrate the efficacy of Quantum Rings' SDK to simulate quantum systems with limited classical compute resources. We show this by executing the RCS circuits published as a part of Google's 2019 work while utilizing 32GB of memory or less and using the XEB to quantify the quality of the execution. In completing this work, we find that the Quantum Rings SDK achieves an average XEB score of 0.678, maintaining a high score even as the circuit complexity increases.

## II.   BACKGROUND

It has been known for some time [1] that constant-depth quantum circuits cannot solve classically intractable decision problems; however, they have remarkable power in sampling probability distributions that cannot be sampled classically in polynomial time. Therefore, if we have to demonstrate quantum advantage, one way is to shift our attention from decision and function problems to sampling problems: that is, problems where the goal is to sample an $n$-bit string, either exactly or approximately, from a desired probability distribution. Furthermore, beyond the quantum advantage experiments, sampling problems can serve as an efficient tool to benchmark quantum systems, which is the primary purpose of our study.

Uniform random bits are binary values (0s and 1s) generated such that each bit has an equal probability of being 0 or 1. A sampling algorithm transforms such uniform random bits into non-uniformly distributed random bits. By running the algorithm many times, we can generate a statistical sample that can provide meaningful insights into the underlying problem [2]. Several such sampling algorithms have been published in the literature. For example, boson sampling involves arranging $n$ bosons in an input formation $k$ and scattering them via a passive linear unitary transformation $U$ into $m >> n$ output modes. The boson sampling problem is to produce a fair sample of the output probability distribution $P(l|k, U)$, where $l$ is the output arrangement [2]. Due to the random nature of $U$, the problem lies in the complexity space[1] #P, making it difficult to simulate. Another statistical model for random sampling from a target probability distribution is the Markov chain Monte Carlo (MCMC) approach. Markov chains start from an initial state and repeatedly jump to new states according to a transition rule. The Metropolis-Hastings algorithm uses a similar approach. Markov chains can help estimate statistics of the target distribution in systems where probability distributions are inaccessible. Besides, Markov chains are used for sampling from Boltzmann distributions and hence have applications in solving combinatorial optimization problems using simulated annealing. However, sampling from the Boltzmann distributions is a hard problem because of the exponential number of parameters. Quantum Monte Carlo Methods (QMC) can calculate the quantum mechanical properties of a system through approximation with Monte Carlo sampling. Boltzmann machines have another application: machine learning. The Boltzmann distribution of the parameterized Ising energy in a spin-glass is a learnable probability distribution over the parametrized discrete domain. Random circuit sampling (RCS) is another method that uses a sequence of randomly chosen gates, creating a complex, highly entangled quantum state. The probabilities of the output states of the random circuits follow a Porter-Thomas distribution, where most bitstrings have very low probabilities, and a few have relatively higher probabilities, a characteristic of chaotic quantum systems.

The question is, if we were to use random circuits, how do we verify the output of the random circuit sampling? If we are able to quantify this, then random circuits can form the metric for benchmarking quantum systems – both physical and simulated. One method to verify the correctness of the quantum output distribution is to perform cross-entropy benchmarking (XEB).

---

[1] #P -- The class of function problems of the form $f(x)$, where $f$ is the number of accepting paths of a nondeterministic Turing machine running in polynomial time

This benchmarking involves comparing the experimentally obtained distribution with the ideal theoretical distribution. The cross-entropy difference between the experimental and ideal quantum distributions should be low if the quantum processor is functioning correctly. Additionally, cross-entropy benchmarks are related to fidelity, which we shall establish shortly. Fidelity measures how closely the experimentally obtained distribution matches the expected quantum distribution. High fidelity indicates that the quantum system accurately implements the random circuit. Therefore, the method of random circuit sampling can be an excellent choice to benchmark the performance of a physical quantum processor or a quantum simulator.

Random circuit sampling uses a quantum circuit defined by a unitary matrix $U$ of size polynomial in $n$ ( $n$ = the total number of qubits) and applies it to an initial state $|\psi_0\rangle$. The unitary matrix $U$ is chosen uniformly from the *Haar measure*[2]. The output state $|\psi\rangle$ is sampled $k$ times, producing bitstrings $\{x_1, x_2, ..., x_k\}$ from the distribution defined by the probabilities of the output state $p(x_i) = |\langle x_i|\psi\rangle|^2$. Each bitstring has a width $n$. Since we are dealing with a quantum system, the expectation value of the bitstrings is not equal to $\frac{1}{N}$ (if it were, it would be a classical uniform sampling), where $N = 2^n$ is the dimensionality of the corresponding Hilbert space, that is $\langle p(x_i)\rangle ! = 2^{-n}$. It has been observed that this probability follows a Porter-Thomas distribution, outlined by the following relation:

$$\mathcal{P}(p) = N\, e^{-Np}, \qquad (1)$$

where $\mathcal{P}(p)$ denotes the probability density function over $p = p(x) = |\langle x|\psi\rangle|^2$ for some state $|\psi\rangle$ obtained from the random quantum circuit.

If there is an ideal quantum machine with a high fidelity, the linear cross-entropy difference between the ideal system and the experimental system is given by the following relation:

$$\mathcal{F}_{XEB} = N\left(\frac{1}{k}\sum_{i=1}^{k}p(x_i)\right) - 1, \qquad (2)$$

where $k$ is the number of samples.

Suppose the experimental system is noiseless, then $\mathcal{F}_{XEB} \to 1$. In a totally noisy environment or if the sampling is uniform, then $\mathcal{F}_{XEB} \to 0$. Since there is a correlation between cross-entropy benchmarking and fidelity, the cross-entropy difference serves as a good indicator of the performance of the experimental system. In addition to benchmarking quantum systems, cross-entropy benchmarking is also used in the calibration of single- and two-qubit gates.

There is sufficient evidence in literature that XEB has a good correspondence with the fidelity. An XEB procedure [4] uses a set of $m$ cycles of random quantum circuits $\mathcal{U} = \{U_1, U_2, ..., U_S\}$ on an experimental quantum system with $n$ qubits. Each circuit is executed $k$ times on the quantum system, producing a set $\mathcal{B}$ of $S \cdot k$ bitstrings $x_{i,j}$ sampled from the distribution $p_{exp}(x_{i,j}) = \langle x_{i,j}|\rho_j|x_{i,j}\rangle$, where $\rho_j$ is the output state of the experiment with circuit $U_j$.

We then use equation (2) to estimate $\mathcal{F}_{XEB}(\mathcal{B}, \mathcal{U})$ averaged over circuits $\mathcal{U}$, with fidelity $F = \langle\psi_j|\rho_j|\psi_j\rangle$, where $|\psi_j\rangle$ is the ideal target state and $\rho_j$ is the output state of the experiment. The result quantifies how well the experimental quantum system can realize quantum circuits of size $n$ and depth $m$.

Artu et al. [5] observe that if an error gate $E$ is inserted at a particular location in $U_j$, the probability $p$ that no error other than $E$ occurs is approximately equal to the experimental fidelity $F$, approximated by $\mathcal{F}_{XEB}(\mathcal{B}, \mathcal{U})$:

$$\mathcal{F}_{XEB}(\mathcal{B}, \mathcal{U}_E) \approx e\, \mathcal{F}_{XEB}(\mathcal{B}, \mathcal{U}), \qquad (3)$$

where $e$ is the probability that $E$ occurs and $\mathcal{U}_E = \{U_{1,E}, U_{2,E}, ..., U_{S,E}\}$ are the quantum circuits obtained from $U_j$ by inserting an error gate $E$ at a particular location.

---

[2] See [3] for an explanation of Haar measure.

From equation (3), we can infer that the XEB result obtained using $\mathcal{U}_E$ is approximately proportional to the XEB result obtained using the error-free reference circuits $\mathcal{U}$; therefore, this allows us to estimate the median probability of a Pauli error. Further, Artu et al. found that more than one gate failure can manifest as a Pauli error $E$ at a particular circuit location, and the measured XEB error agrees well with the single- and two-qubit randomized benchmarks.

In a system with homogeneously distributed random errors at the rate $\epsilon$, the probability that the circuit executes without error is $p = (1 - \epsilon)^{\#gates}$. From equation (3), we find that a single or more error can significantly reduce $\mathcal{F}_{XEB}$. Therefore, we can assume that in a low-error regime, the fidelity and $\mathcal{F}_{XEB}$ approximately correspond to no error probability.

$$F \approx \mathcal{F}_{XEB}(\mathcal{B}, \mathcal{U}) \approx (1 - \epsilon)^{\#gates}, \quad (4)$$

Hence, a measure of $\mathcal{F}_{XEB}$ suggests the fidelity of the experimental quantum system, especially if the sampled distribution is in the Porter-Thomas regime, given by equation (1). In accordance with these observations, we shall use the cross-entropy benchmarking to study the performance characteristics of the Quantum Rings SDK.

## III. THE EXPERIMENT

Our experiments utilized the standard gate sequence published by Google in the dataset [7] dated June 13, 2022. The gate sequence employed by Google is shown in Figure 1. Each cycle of the algorithm consists of applying single-qubit gates chosen randomly from the set $\{\sqrt{X}, \sqrt{Y}, \sqrt{W}\}$, where $W = \frac{X+Y}{\sqrt{2}}$, on all qubits, followed by two-qubit gates on pairs of qubits. The 2-qubit interaction gates use Google's implementation of a Fermionic simulation gate (FSimGate)[3]. The 2-qubit gates are surrounded by a set of single qubit rotations. Each random circuit used $m$ cycles of the gate sequences. The source code of the random circuits can be downloaded from Google's Dryad repository [7].

Among the six types of random circuits provided by Google, we selected the full circuits ($e0$) with the number of cycles set to $m = 14$, pattern $p = EFGH$, and the following configuration:

- n: number of qubits
  $\{12,14,\ldots 38,39,\ldots 51,53\}$
- s: PRNG seed number $\{0,1,\ldots,9\}$

The pattern $p = EFGH$ indicates that a layer of random 1-qubit gates is initially applied to all qubits, followed by the application of 2-qubit gates according to $E$. This pattern is repeated with patterns $F$, $G$, and $H$ respectively.

For all of the circuits, Google provided QASM files, which we loaded directly using the Quantum Rings SDK to create the equivalent quantum circuit. Google also provided the amplitudes for most of these circuits, which we used to calculate the $F_{XEB}$ for comparison. Except for the circuits with $n = \{51, 53\}$, we sampled the remaining circuits 500,000 times each. Each of the circuits with $n = \{51, 53\}$ was sampled 2,500,000 times. This was done to ensure that the number of samples matched those in Google's amplitude files. All samples obtained from the circuits were used for probability calculations.

## IV. QUANTUM RINGS SDK

The Quantum Rings SDK [20] is a comprehensive quantum development platform compatible with Google Colab, Linux, macOS, and Windows operating systems. The SDK supports exporting and importing QASM files, ensuring compatibility with existing quantum codebases. An accompanying toolkit enables the execution of Qiskit-developed quantum circuits directly on the Quantum Rings backend. Central to the toolkit is a simulator capable of executing

---

[3] "Cirq.FSimGate | Cirq." n.d. Google Quantum AI. Accessed July 11, 2024. https://quantumai.google/reference/python/cirq/FSimGate.

all well-known single--, two--, and multi-qubit gate operations.

The circuits were executed on two distinct system configurations. The first utilized a Dell Alienware R15 system equipped with Intel(R) Core(TM) i9-14900KF, 3200 Mhz, 24 Core(s), 32 Logical Processor(s), 32GB of memory, and Windows 11 Pro. The second configuration involved a high-performance node on the Sol supercomputer [17] at Arizona State University, featuring 32 AMD EPYC 7713 64-core Processors with 32 GB of memory each, running Rocky Linux 8.9 (Green Obsidian).

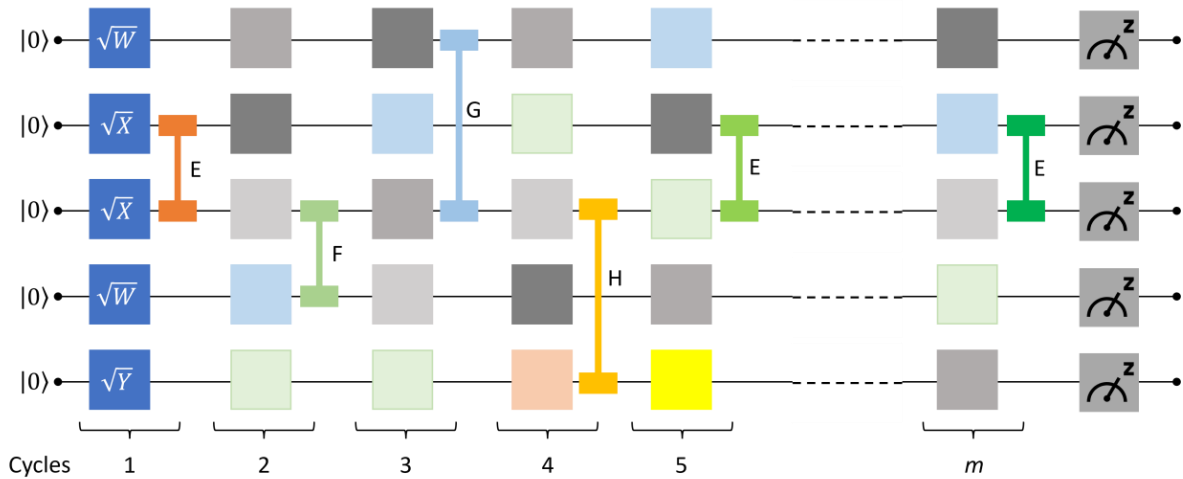

**Figure 1. An example of the random circuit instance used by Google.**

Executing the circuits using the Quantum Rings SDK is straightforward. We first downloaded Google's Dryad repository, dated June 23, 2022 [7], into a local folder and unzipped all the '*tar.gz*' files. We retained only the essential QASM and amplitude files and discarded the rest. The Quantum Rings SDK was then installed using the installation procedure outlined in the SDK documentation[4]. Utilizing the source code from Quantum Rings' public repository [19], we executed the random circuits (contained in the QASM files downloaded from the Dryad repository) in the target systems in the Python environment. The source code also includes the routines to plot the graphs this paper illustrates.

For all the circuits, the SDK was used to output the amplitudes of each measurement, from which the corresponding probabilities were calculated by taking the square of the absolute value of the amplitude. While the measured probabilities uniformly explore the Hilbert space, certain states are observed to be favored, like that of a laser's speckle pattern, as illustrated in Figure 2. This distribution is given by $e^{-Np}$, which is the well-known Porter-Thomas distribution. Our experiments revealed that the probabilities of the measurements closely align with the Porter-Thomas distribution, consistent with the literature [8], thereby illustrating the underlying quantum dynamics.

The Shannon entropy of a set of probabilities is logarithmic, and it is given by $H(P) = -\sum_i p_i \log(p_i)$. The cross-entropy between two sets of probabilities is then $H(P, Q) = -\sum_i p_i \log(q_i)$. The authors of Google's main paper [4] used linear terms instead of the logarithmic form, a choice that is gaining traction due to its computational efficiency. This approach is now becoming the standard in academia and industry. We performed linear XEB calculations for all the circuits using equation (2), adhering to this prevailing norm. As outlined in literature, noiseless simulation results in $F_{XEB} = 1$, and uniform samples results in $F_{XEB} = 0$. A non-vanishing value of $F_{XEB}$ means that the sampled distribution correlates with the ideal one. The experimental results showed a strong correlation with ideal simulation. Figure 4 illustrates the linear XEB we obtained from the Quantum Rings SDK, and the values calculated from Google's amplitude files for side-by-side comparison.

We then measured the actual time required to obtain the first sample of the largest full circuits ($e0$) with $n = 53$, $m = \{12,14,16,18,20\}$, $p = "ABCDCDAB"$. Our observations indicated that the largest circuits were executed within a reasonable amount of timeframe. Figure 5 illustrates our execution time measurements.

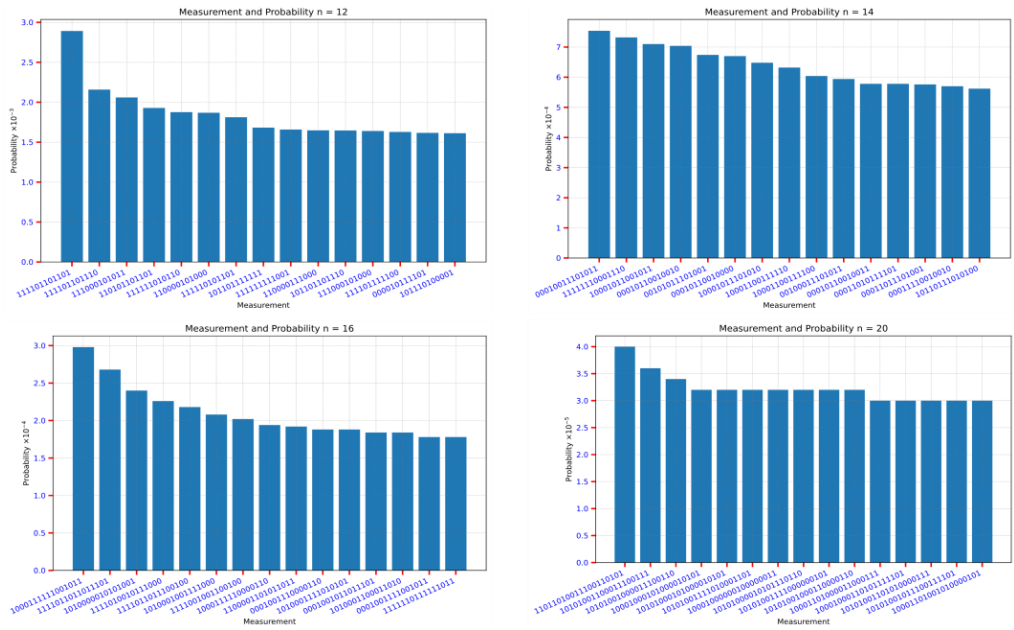

**Figure 2. Measurement probabilities**. This graph gives significant insights into the fact that some states are favored and follow a distribution. Only the top 10 states are shown.

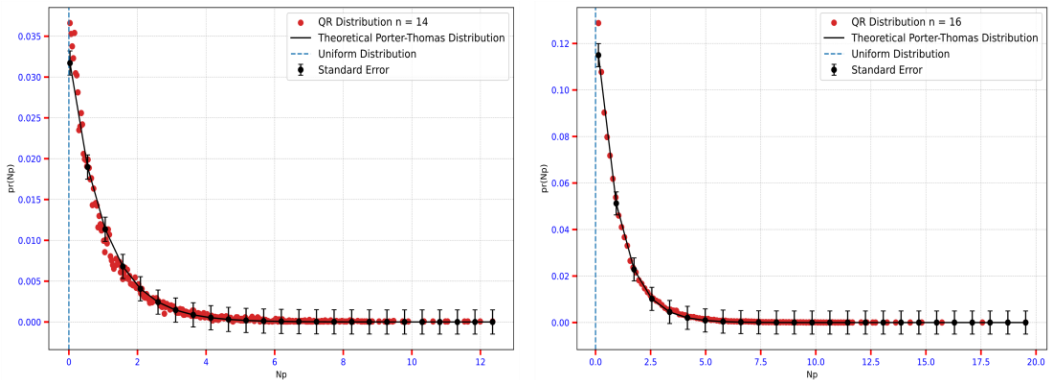

**Figure 3. Plots illustrating the Porter-Thomas distribution of the samples.** The blue line shows the uniform distribution. The black line shows the theoretical Porter-Thomas distribution. The red dots are the samples from the Quantum Rings SDK. The Quantum Rings distribution converges with the theoretical Porter-Thomas distribution within an excellent error margin, showing the underlying quantum dynamics.

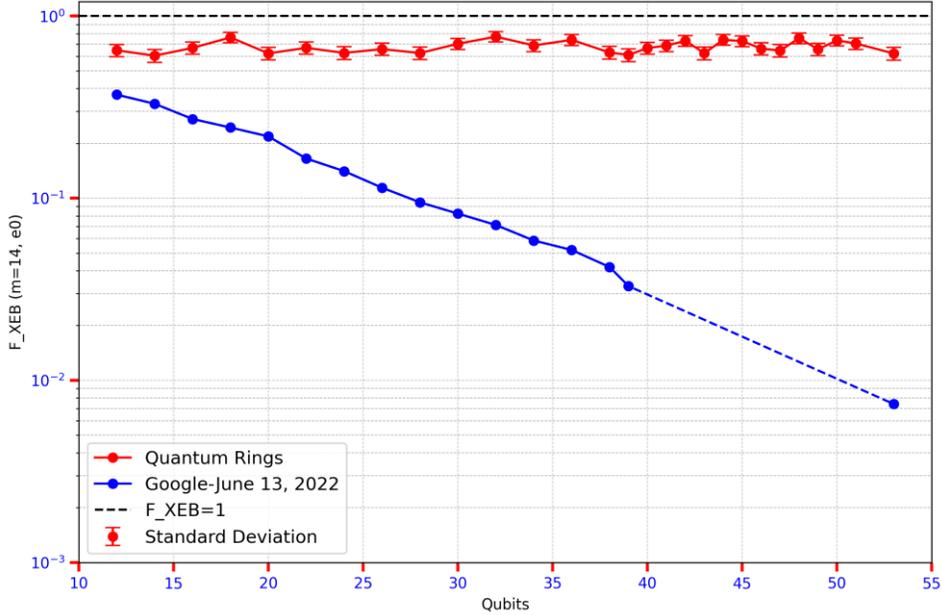

**Figure 4. Graph illustrating the $\mathcal{F}_{XEB}$ for circuits with $m = 14$, $e0$, $p = 'EFGH'$.** The blue line shows the values we calculated from the corresponding Google amplitude files. The dashed segment is the qubits range for which Google's amplitudes are unavailable. The red colored line shows the values calculated using the Quantum Rings SDK. The average $\mathcal{F}_{XEB}$ obtained was 0.678. We obtained $\mathcal{F}_{XEB} = 0.622$ at 2.5 million samples for the largest circuit with $n = 53$. The error bars correspond to the standard deviation of the samples.

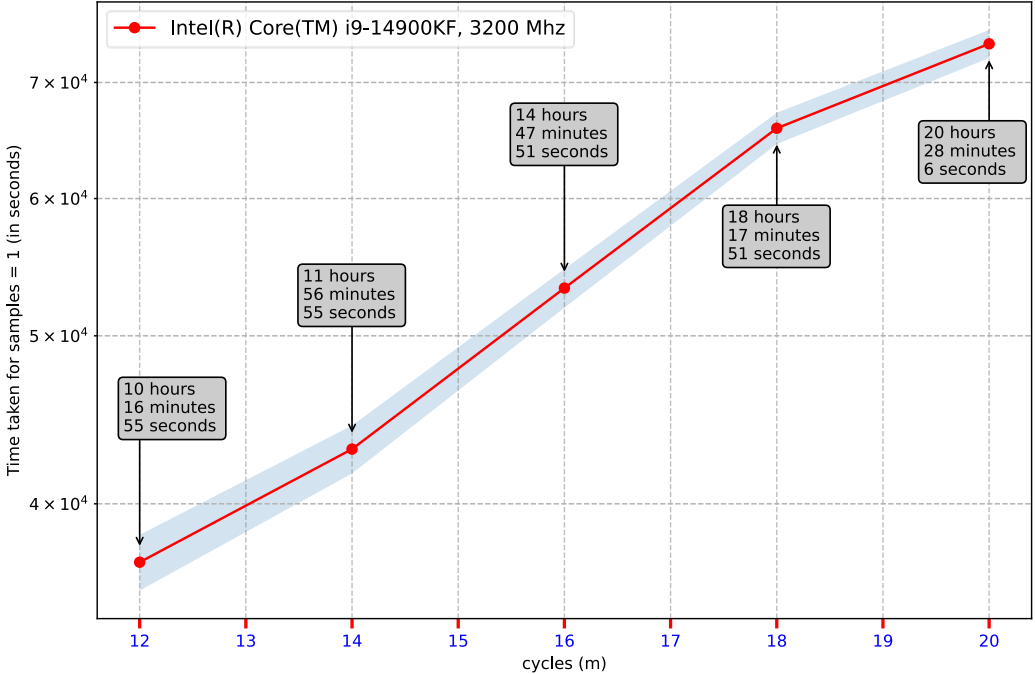

**Figure 5. Circuit time analysis.** This graph shows the time taken to obtain the first sample of full circuits with $n = 53$ $m = 12,14,16,18,20$, $e0$, and $p = "ABCDCDAB"$.

## V. CONCLUSION

Our experiments demonstrate that the Quantum Rings SDK is capable of effectively simulating Google's Sycamore circuits, producing random circuit samples whose probabilities are consistent with the Porter-Thomas distribution, thereby illustrating quantum dynamics.

The average linear cross-entropy benchmarking (XEB) score of 0.678 indicates a strong correlation with ideal quantum simulation and represents the highest fidelity observed to date [4-7,9,10,12,15]. Execution times validate that such simulations can be run efficiently on standard developer hardware.

This capability enables researchers and developers to build, debug, and execute large-scale quantum algorithms and applications, providing a critical developer capability in anticipation of future advancements in physical quantum computing. Nevertheless, this advancement does not diminish the need for continued development of physical quantum computers. The advancements achieved underscore the importance of relentless efforts in both simulation and hardware development to realize practical quantum computing solutions.

## ACKNOWLEDGEMENTS

The Quantum Rings Team would like to thank:

- Dr. John Martinis for his insights into the need for XEB analysis to validate the SDK, which inspired this paper.
- Dr. Gil Speyer at Arizona State University for providing access to the computing resources (Sol) required for this experiment [17].
- The National Center for Supercomputing Applications (NCSA) at the University of Illinois Urbana-Champaign for replicating the experiment using the Delta supercomputer, supported by NSF Grant 2005572. We would like to thank Dr. Santiago Núñez-Corrales for facilitating this collaboration, Darren Adams for executing the experiment, and the extended NCSA team for their support.
- The Quantum Collaborative at ASU

## DATA AVAILABILITY

The source code and the amplitude files for all circuits are available to the public at [19]. The Quantum Rings SDK can be downloaded from [20].

---

[20] The Quantum Rings SDK can be installed by following the instructions at: https://quantumrings.com/doc/index.html.

License keys can be obtained by registering at https://www.quantumrings.com.
(free for academic and non-commercial use)